\pdfoutput=1
\documentclass[journal=jacsat,manuscript=article]{achemso}

\usepackage{chemformula} 
\usepackage{graphicx}%
\usepackage{multirow}%
\usepackage{amsmath,amssymb,amsfonts}%
\usepackage{amsthm}%
\usepackage{mathrsfs}%
\usepackage[title]{appendix}%
\usepackage{xcolor}%
\usepackage{textcomp}%
\usepackage{manyfoot}%
\usepackage{booktabs}%
\usepackage{algorithm}%
\usepackage{algorithmicx}%
\usepackage{algpseudocode}%
\usepackage{listings}%

\usepackage{siunitx}
\usepackage{float}
\usepackage[normalem]{ulem}
\usepackage{rotating} 
\usepackage{amssymb}
\usepackage{bibentry}

\author{Daria Tatsii}
\affiliation{Department of Meteorology and Geophysics, University of Vienna, Universitätsring 1, 1010, Vienna, Austria}
\email{daria.tatsii@univie.ac.at}
\author{Silvia Bucci}
\affiliation{Department of Meteorology and Geophysics, University of Vienna, Universitätsring 1, 1010, Vienna, Austria}

\author{Taraprasad Bhowmick}
\affiliation{Laboratory for Fluid Physics, Pattern Formation and Biocomplexity, Max Planck Institute for Dynamics and Self-Organisation, Am Faßberg 17, 37077, G\"ottingen, Germany}
\alsoaffiliation{Institute for the Dynamics of Complex Systems, University of G\"ottingen, Friedrich-Hund-Platz 1, 37077, G\"ottingen, Germany}

\author{Johannes Guettler}
\affiliation{Laboratory for Fluid Physics, Pattern Formation and Biocomplexity, Max Planck Institute for Dynamics and Self-Organisation, Am Faßberg 17, 37077, G\"ottingen, Germany}

\author{Lucie Bakels}
\affiliation{Department of Meteorology and Geophysics, University of Vienna, Universitätsring 1, 1010, Vienna, Austria}

\author{Gholamhossein Bagheri}
\affiliation{Laboratory for Fluid Physics, Pattern Formation and Biocomplexity, Max Planck Institute for Dynamics and Self-Organisation, Am Faßberg 17, 37077, G\"ottingen, Germany}
\email{gholamhossein.bagheri@ds.mpg.de}
\altaffiliation{Contributed equally to this work}

\author{Andreas Stohl}
\affiliation{Department of Meteorology and Geophysics, University of Vienna, Universitätsring 1, 1010, Vienna, Austria}
\altaffiliation{Contributed equally to this work}

\title{Shape matters: long-range transport of microplastic fibers in the atmosphere}

\keywords{microplastics $|$ shape $|$ atmospheric transport $|$ gravitational settling}

\begin{document}

\begin{abstract}
Deposition of giant microplastic particles from the atmosphere has been observed in the most remote places on Earth. However, their deposition patterns are difficult to reproduce using current atmospheric transport models. These models usually treat particles as perfect spheres, whereas the real shapes of microplastic particles are often far from spherical. Such particles experience lower settling velocities compared to volume-equivalent spheres, leading to longer atmospheric transport.
Here, we present novel laboratory experiments on the gravitational settling of microplastic fibers in air and find that their settling velocities are reduced by up to 76$\%$ compared to spheres of the same volume.
An atmospheric transport model constrained with the experimental data shows that shape-corrected settling velocities significantly increase the horizontal and vertical transport of particles.
Our model results show that microplastic fibers of about $\SI{1}{\milli\meter}$ length emitted in populated areas can reach extremely remote regions of the globe, including the High Arctic, which is not the case for spheres. We also calculate that fibers with lengths of up to $\SI{100}{\micro\meter}$ settle slowly enough to be lifted high into the stratosphere, where degradation by ultraviolet radiation may release chlorine and bromine, thus potentially damaging the stratospheric ozone layer. These findings suggest that the growing environmental burden and still increasing emissions of plastics pose multiple threats to life on Earth.
\end{abstract}

\section{Introduction}
Airborne particles have a large impact on Earth's energy balance, atmospheric composition, water and carbon cycles, ecosystems, and human health \cite{SeinfeldPandis2016}.
Therefore, it is crucial to know their distribution in the global atmosphere, for which an accurate understanding of their emissions and removal mechanisms is essential. 
Accumulation mode particles sized 100-1000~nm are known to be transported over long distances in the atmosphere\cite{Matsui2011}, sometimes across an entire hemisphere \cite{Damoah2004}. 
However, several studies \cite{Varga2021,Does2018,Jeong2013,Madonna2010,MIDDLETON2001411,Betzer1988} have found that even "giant" particles larger than $\SI{75}{\micro\meter}$ in diameter can stay airborne for extended periods and be deposited thousands of kilometers away from their sources.
This is difficult to reproduce with current atmospheric transport models, which predict a much shorter dispersion range for such large particles \cite{ZHANG20191,Liu2014}.
The question is what mechanisms or model deficiencies can explain these discrepancies.

Several mechanisms that have been suggested to enhance transport distances, for instance, strong winds \cite{Knippertz2011}, strong turbulence keeping individual particles aloft \cite{Garcia-Carreras2015}, or electric forces counteracting the particles' weight \cite{Nicoll2012,Renard2018} seem insufficient to resolve the model problems.
Another possibility investigated in this paper is the effect of particle shape.
Most regional or global transport models treat particles as perfect spheres, whereas in reality, their shapes are often far from spherical. Such particles experience a larger drag in the atmosphere compared to spheres, which reduces their settling velocity and facilitates longer transport distances \cite{SAXBY2018}.
This is particularly true for microplastics, which are often found as fibers or with other complex shapes in the environment and also have a lower density than many other aerosol types \cite{BERGMANN2019,ALLEN2019,AMBROSINI2019,Brahney2021,Stefansson2021,CABRERA2020,GONZALEZPLEITER2020}.

To date, shape corrections for gravitational settling calculations\cite{BB2016} have been mostly used for modeling dispersion of volcanic ash \cite{SAXBY2018} and mineral dust \cite{Mallios2020}.
However, microplastic fibers with lengths that exceed their thickness by a factor of 40 or more, represent a greater challenge.
While their settling behavior in water or other liquids is already relatively well-constrained, 
only a few experiments have investigated the settling of non-spherical particles in air \cite{Bhowmick2023,BB2016,Newsom1994,JayaweeraCottis1969} and the available data do not cover the range of sizes and shapes relevant to microplastics.
In particular, data on the settling behavior of fibers, especially of bent shapes, are missing and this limits our capability to reliably simulate the dispersion of microplastics in the atmosphere.

To determine the settling velocities of microplastic fibers of different sizes and shapes, we performed a series of experiments in a newly developed setup (SI Figure S1a) \cite{Bhowmick2023}.
Particles with a mass density of $\SI{1200}{\kg\per\meter\cubed}$, typical of common plastic polymers, and of various shapes including spheres, straight fibers, and curved fibers with semi- and quarter-circular shapes with aspect ratios AR (the ratio of the fiber length to its diameter) of 20, 50, and 100 were produced using 3D printing technology \cite{NanoScribe_2017} (Figure 1a). Detailed information about the experiments can be found in Methods.

\begin{figure*}
\centering
\includegraphics[width=0.8\textwidth]{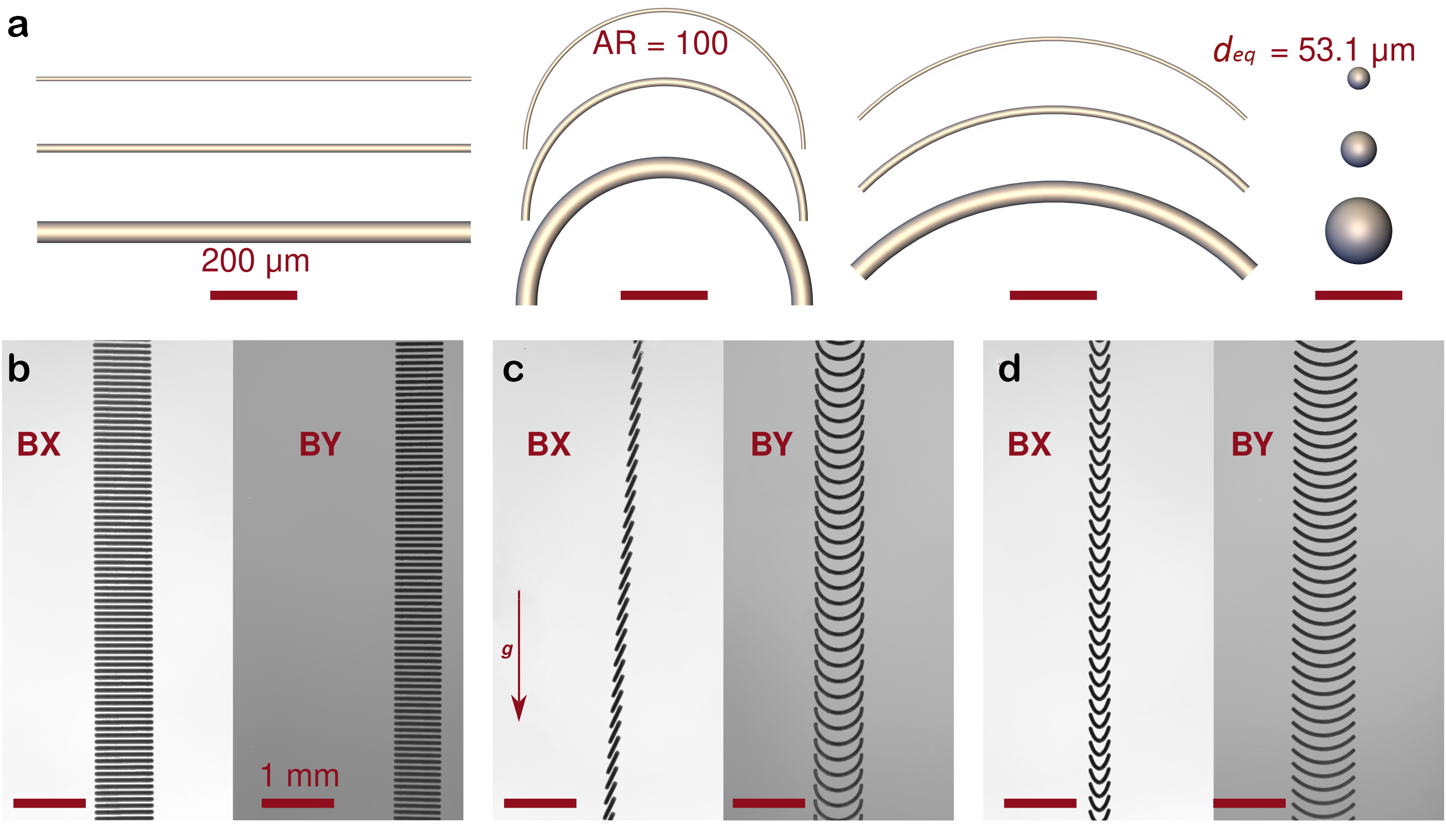}
\caption{Example of the particles printed for the experiments and settling behavior of fibers.(\textbf{a}) Straight, semicircular, and quarter circular \SI{1}{\milli\meter} fibers for different aspect ratios AR, and the corresponding diameters $d_{eq}$ of volume equivalent spheres (sizes of other printed particles are listed in SI Table S1). 
Snapshots of the \SI{1}{\milli\meter} (\textbf{b-d}) fibers with aspect ratio AR $=20$, indicating their settling behavior under gravity.
The snapshots correspond to the last \SI{6.6}{\milli\meter} of recording by the bottom cameras BX and BY from the experiments with (\textbf{b}) straight, (\textbf{c}) semicircular, and (\textbf{d}) quarter circular fibers. BX and BY refer to lower cameras with field of view perpendicular to each other. $g$ shows the gravity vector. $d_{eq}$, i.e. the equivalent-volume diameter, is the diameter of a sphere with the same volume as the particle. }\label{fig:fig1}
\end{figure*}

\section*{Results and Discussion}

Our experimental results reveal that the measured terminal settling velocities for straight fibers, semicircular and quarter circular fibers are only 24-51$\%$ of the velocities of spheres with the same volume (Figure 2a, SI Table S1).
The largest differences are found for the particles with the largest aspect ratios (AR $=100$), for which the settling velocities are less than one-third of those for spheres of the same volume. 
For a given volume and aspect ratio, the differences between the velocities of straight, semicircular, and quarter circular fibers are relatively small (on average, 12$\%$ with a maximum deviation of $\sim$26$\%$ for particles of size of 1000~µm~x~10~µm).
Nevertheless, straight fibers have the lowest and semicircular fibers the highest settling velocities. On average, straight fibers settle with 38$\%$ (minimum value of 24$\%$)  and semicircular fibers with 41$\%$ (minimum value of 29$\%$) of the velocity of spheres of the same volume. In addition to the settling velocities of the fibers, we also studied their orientation dynamics and the interplay of these dynamics in the turbulent atmosphere. Figure S1c-h (Supporting Information) show exemplary super-imposed images from  particles falling through the lower region of the field of view of the bottom cameras, where the fibers had already reached a steady-state velocity.

\begin{figure}
\centering
\includegraphics[width=1\linewidth]{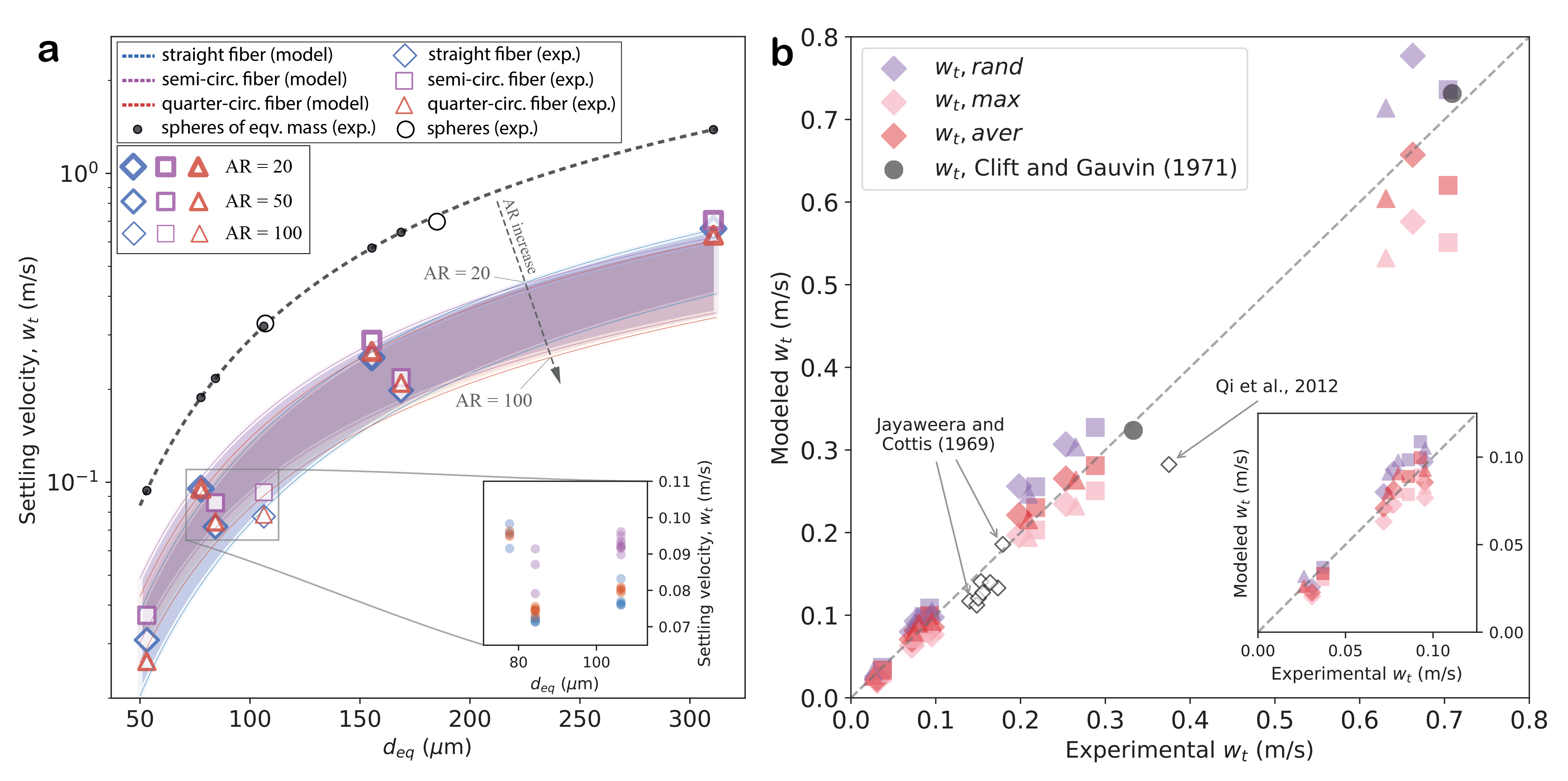}
\caption{Observed and modeled gravitational settling velocities.(\textbf{a}) Settling velocities as a function of particle size, expressed as the diameter of a sphere of equivalent volume. Modeled values using the shape correction scheme for averaged orientation are shown as a black dashed line for spheres and as colored shaded areas covering the range of aspect ratios from 20 to 100 for straight (blue), semicircular (purple) and quarter circular (red) fibers. Experimental mean values for each shape are represented by open markers, coded by color for shape type and by thickness for aspect ratio. The spread in the results of repeated experiments for individual particle types is smaller than the size of the symbols. The inset gives an example of the experimental data spread for a limited size range on a linear velocity scale (semitransparent blue dots for straight fibers, purple dots for semicircular fibers, red dots for quarter circular fibers).
Black dots show the calculated settling velocities of spheres with volumes equivalent to the experimentally studied fibers. (\textbf{b}) Scatter plot of observed and modeled settling velocities. Modeled values were calculated for random (purple), maximum-drag (pink) and averaged (red) orientations. Black open rhombuses show previously published experimental results by Qi \textit{et. al}\cite{QI020121} and Jayaweera and Cottis (1969) \cite{JayaweeraCottis1969}. Particle shapes are distinguished as reported in the legend in panel a. The inset zooms on the smallest settling velocities.}\label{fig:fig2}
\end{figure}

We find in similitude with previous work \cite{Bhowmick2023,QI020121,Newsom1994} that fiber-like particles achieve a terminal stable orientation with their maximum projection area perpendicular to the settling direction. 
Beyond the existing knowledge, we further observe that depending on their aspect ratio, fibers can approach their steady-state orientation with or without oscillations, which is in agreement with the theoretical predictions about oscillations in Figure 4 of Bhowmick \textit{et al} \cite{Bhowmick2023}.
In particular, the fiber oscillations decrease as the fiber diameter decreases (SI Figure S4).   
For example, regardless of their shape, the fibers with \SI{1}{\milli\meter} length and AR $=20$ had reached both their steady-state orientation and settling velocity in the section observed by the bottom cameras (SI Figure S1c-e).
Their oscillations in orientation were already dampened when they fell through the section observed by the top cameras (SI Figure S4a-c).
In contrast, the fibers with \SI{2}{\milli\meter} length and AR $=20$ did not reach a steady-state orientation for the entire duration of the camera recording but they had nevertheless reached a steady-state settling velocity when observed by the bottom cameras (SI Figure S1f-h, Figure S4e-g, and S10).
This is because the settling velocity responds to orientation changes only on time scales longer than the period of orientation oscillations.
In fact, our experiments reveal that the fiber oscillations decay exponentially in air, with a decay constant of the order of \SI{10}{\milli\second}. 
Therefore, the time scales for alignment in fiber orientation are smaller than the smallest time scales typically encountered in atmospheric turbulence.    
An important implication of these new findings is that the orientation of the investigated fibers with respect to the gravity vector is similar in the turbulent atmosphere and in still air.
In situations when these fibers encounter highly turbulent regions of the atmosphere such as in clouds, fiber orientation may become random. 
However, the deviations from the measurements made here would be negligible because (i) these events would be brief compared to the residence time of the fibers in the atmosphere, and (ii) even model calculations for randomly oriented fibers deviate only by about 17$\%$ from our measurements (SI Table S2).

Recently, the model of Bagheri and Bonadonna (2016) \cite{BB2016} was shown to be associated with small errors in estimating the settling velocity of microplastics of various shapes in water \cite{Coyle2023}.
The model can be fitted to predict the settling velocity for non-spherical particles when they fall with their minimum, maximum, or random projection areas facing downward, and it is also applicable to settling in air (see Methods). 
Figure 2b compares this model with the observations.
With its maximum drag orientation configuration, the model agrees well with the measurements but, on average, it systematically underestimates the observed velocities by 13.3$\pm$10.0$\%$ for straight fibers.
For random orientation, the model has slightly higher deviations with respect to the measurements and systematically overestimates them by 16.8$\pm$8.4$\%$.
The better agreement between the maximum projection model and the measurements is consistent with the observation that particles mostly stabilize with their maximum projection area facing downward (SI Figures S1 and S2).
However, even better agreement is found for the averaged orientation, which is the arithmetic mean of the predictions for the maximum and random orientations and allows for some oscillation around the maximum drag orientation.
With the average-orientation model, no systematic bias occurs and the deviations are further reduced to 8.6$\pm$8.2$\%$.
The results for semicircular and quarter circular fibers are similar (SI Table S2).

The excellent agreement between the observed and modeled settling velocities within less than 10$\%$ suggests that we can use the Bagheri and Bonadonna (2016) \cite{BB2016} scheme with averaged orientation to realistically simulate gravitational settling in the atmosphere.
To explore the effects of the shape dependence of gravitational settling on the global transport of microplastics in the atmosphere, we therefore implemented this scheme into the Lagrangian atmospheric transport model, FLEXPART \cite{Stohl2005, Pisso2019} (SI Table S3).

With the extended model, we performed 1-year simulations of particle transport in the atmosphere for spheres as well as for straight fibers with aspect ratios of 20, 50, and 100.
All particles had the same volume, corresponding to an equivalent sphere with a diameter of $\SI{75}{\micro\meter}$, which can be considered as giant particles \cite{Does2018, Jeong2013}.
We ran two types of simulations:
in the first set of simulations, we released particles in instantaneous pulses 10-100~m above the ground once per day at local noon from five release points to gather statistical information about their atmospheric lifetimes and transport distances.
In the second set of simulations, we used a realistic scenario of land-based microplastic emissions and explored the resulting global atmospheric concentration and deposition patterns for spheres and fibers (see Methods).

The deposition patterns resulting from the first set of simulations clearly show that the atmospheric transport distances depend systematically on the particle shape (Figure 2 and SI Figure S5). 
For instance, for a release point in northern Italy, for spheres virtually all deposition occurs in central and southern Europe, whereas straight fibers with AR $=100$ are also deposited in northern Africa and northern Europe and small amounts even reach the Arctic (SI Figure S6a-d).

\begin{figure*}
\centering
\includegraphics[width=0.9\textwidth]{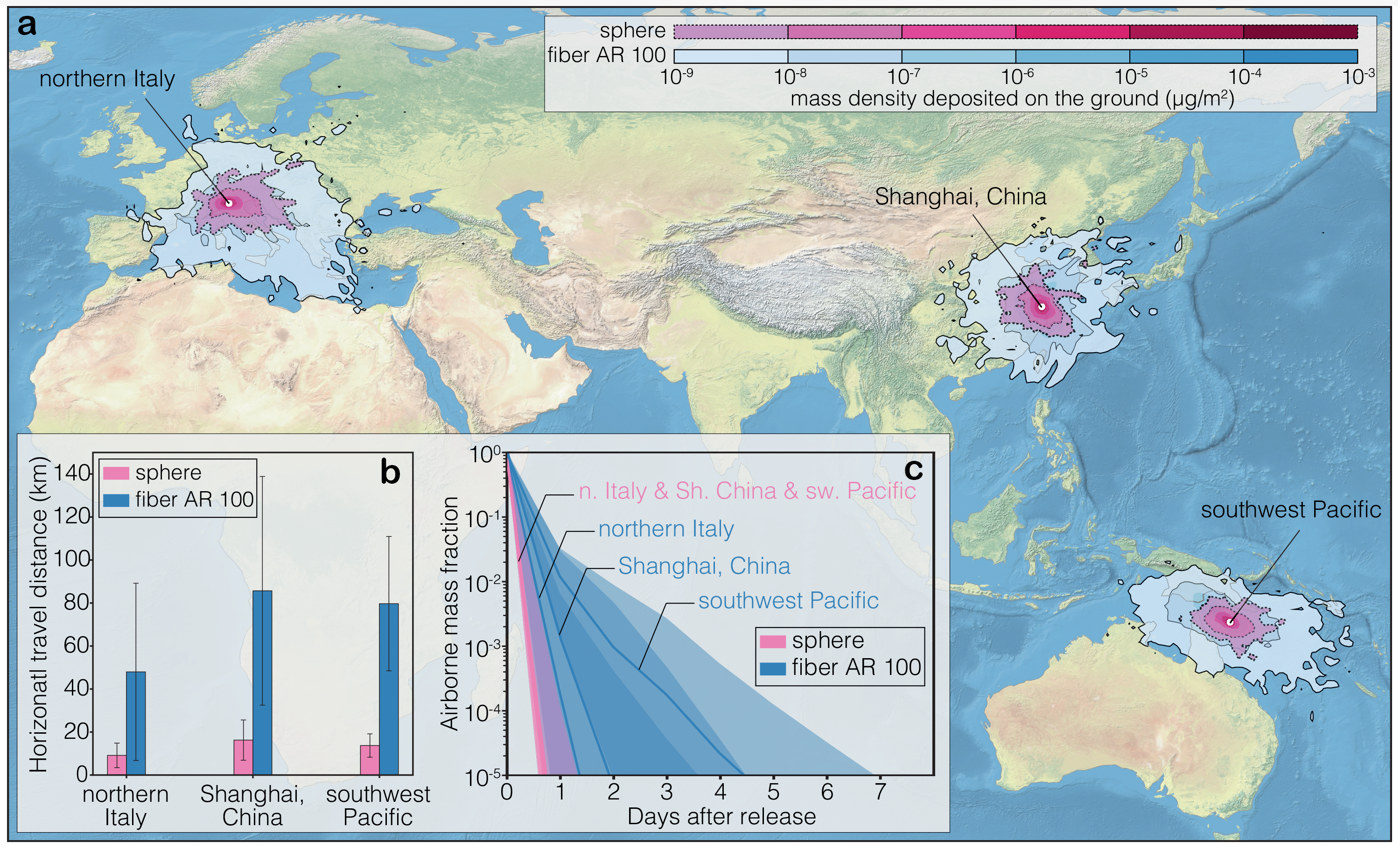}
\caption{Characteristics of microplastics transport in the atmosphere. Shown are FLEXPART model simulation results for three release points (northern Italy; Shanghai, China; southwest Pacific) and for microplastic particles with an equivalent diameter of $\SI{75}{\micro\meter}$ for spheres (pink) and straight fibers with an aspect ratio of 100 (blue). (\textbf{a}) Shown on the map is the annual mean total deposition from the atmosphere for spheres and fibers. (\textbf{b}) Annual mean values of the horizontal transport distances (colored bars) and their standard deviation (whiskers) for spheres and fibers released from the three points. (\textbf{c}) Decrease of the atmospheric microplastic burden as a function of time after release of spheres and fibers from the three points, with the solid lines showing the median values and the shading indicating the range between the 25th and 75th percentiles.}\label{fig:fig3}
\end{figure*}

On average across our five release points, the mean atmospheric transport distances of fibers with aspect ratios of 20, 50, and 100 are, respectively, 157$\pm$26\%, 272$\pm$50\%, and 394$\pm$79\% greater than for spheres with the same volume (SI Figure S5).
This is a result of the longer residence time of fibers in the atmosphere compared to spheres, with e-folding times of 2.4, 4.1, 4.6, and 5.3 hours for spheres and straight fibers with aspect ratios of 20, 50, and 100, respectively.
Consequently, the median fractions of their emitted mass residing in the atmosphere differ by orders of magnitude after a few days (Figure 3c and SI Figure S7). 
The spread in the mass fraction frequency distributions is also broader for fibers than for spheres, indicating a strongly enhanced probability for very long range transport.
This explains why deposition and mass concentration fields extend over much larger regions for fibers than for spheres (Figure 3a and SI Figures S6 and S8).

The second set of simulations shows that the deposition of spheres is strongly focused on the densely populated source regions, while fibers are deposited globally (Figure 4a,b and SI Figure S9a-d).
For the fibers with AR $=100$, only remote regions in the south Pacific and interior Antarctica remain nearly unaffected by microplastic deposition.

For example, our simulations show a total deposition of 2.4~t/year in the high Arctic (north of 75°N) for fibers with AR $=100$, while volume equivalent spheres do not reach this region at all (SI Table S4). 
Our results explain why the high Arctic, despite its remoteness, has been reported to be heavily affected by microplastic deposition from the atmosphere \cite{HAMILTON2021,BERGMANN2019}, and why most of the microplastic particles found there appear to be fibers\cite{HAMILTON2021}.

Many studies discuss the impact of microplastic on the World Ocean \cite{Lebreton2017,Jambeck2015,ANDRADY2011}, however, the contribution of atmospheric transport to microplastic ocean contamination remains unknown.
Our simulations show that if all particles are shaped like spheres, 16~kt/year are deposited in the oceans, whereas for straight fibers with aspect ratios of 20, 50, and 100, the corresponding numbers are 18, 20, and 23~kt/year, respectively. 
Furthermore, spheres are deposited mainly in coastal regions close to population centers, whereas the straight fibers also reach remote ocean areas.
For the remote central Indian Ocean (SI Figure S2), $\sim$7 times more microplastic mass would be deposited if particles are shaped as fibers with AR $=100$ than if they are spheres (SI Table S4).

Remote land regions are also contaminated more severely for fibers than for spheres, as shown in Figure 4a,b.
For instance, the interior of Australia (SI Figure S2) receives 90$\%$ more microplastic mass for fibers with AR$=100$ than for spheres.
These results are in good agreement with the fact that in relatively remote regions, such as national parks, mainly microplastic fibers and other non-spherical particles have been reported \cite{Brahney2021,Stefansson2021,CABRERA2020,GONZALEZPLEITER2020,ALLEN2019,AMBROSINI2019,BERGMANN2019}, whereas reports of microplastic spheres are less common.

The simulated vertical distribution of microplastic particles in the atmosphere is also very different for spheres and straight fibers (Figure 4c,d), with the fibers reaching much higher altitudes than spheres.
For instance, the microplastic mass at altitudes greater than 4~km above the surface is 5.5, 9.0, and 13.3 times higher for straight fibers with AR $=20$, AR $=50$, and AR $=100$, respectively, than for spheres (SI Figure S9e-h).
The effective transport of fibers to high altitudes could have substantial implications for ice cloud formation, since microplastic particles are thought to serve as ice nuclei \cite{Ganguly2019}.
The larger surface area of fibers compared to that of spheres of the same volume is another important factor, possibly making microplastic fibers highly effective ice nuclei with an impact on climate.

\begin{figure*}
\centering
\includegraphics[width=0.9\textwidth]{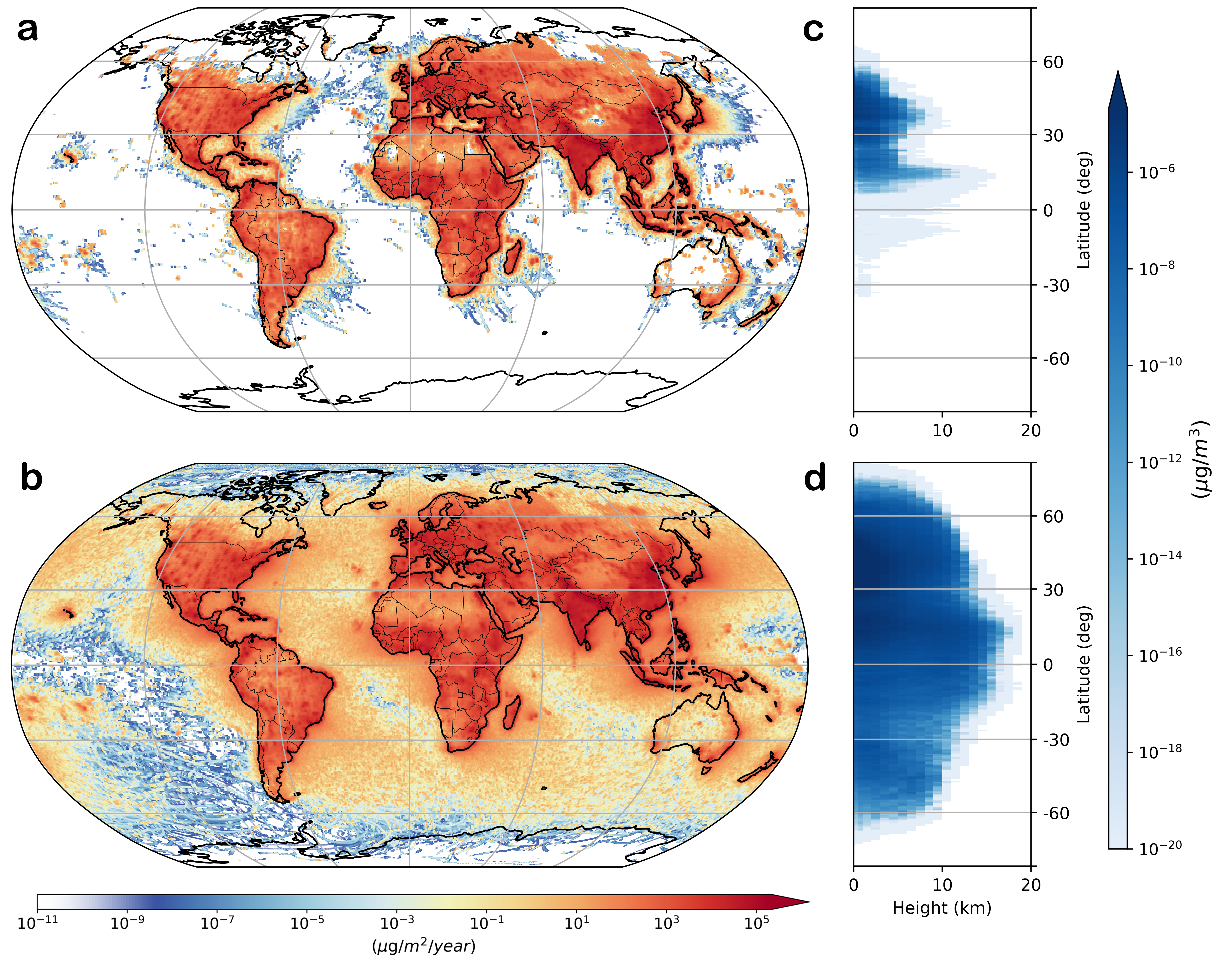}
\caption{Shape dependence of global microplastic deposition and vertical distribution in the atmosphere. Shown are results for particles with an equivalent diameter of $\SI{75}{\micro\meter}$ that only deviate in their shape, for (\textbf{a}) spheres and (\textbf{b}) straight fibers with an aspect ratio of 100. Zonal median atmospheric mass concentration of microplastic as a function of latitude and altitude for (\textbf{c}) spheres and (\textbf{d}) straight fibers with an aspect ratio of 100.}\label{fig:fig4}
\end{figure*}

It seems even likely that microplastic fibers of smaller sizes can reach the stratosphere, with potentially severe consequences for the ozone layer.
For fibers with AR $=100$ and a density of $\SI{900}{\kg\per\meter\cubed}$, typical of polyurethane (density of $\SI{1400}{\kg\per\meter\cubed}$, typical of polyvinyl chloride), we obtain settling velocities of less than 0.2~mm/s at stratospheric altitudes between 20 and 30~km for fiber lengths of $\SI{94}{\micro\meter}$ ($\SI{75}{\micro\meter}$).
This is slower than the minimum mean upwelling velocity of the Brewer-Dobson circulation near the tropical tropopause \cite{Diallo2021,Flury2013}, leading to the fibers' ascent into the stratosphere, where removal processes such as gravitational settling, chemical degradation or  incorporation into polar stratospheric clouds are not effective and can take years for the particle sizes mentioned.
Many plastic products contain bromine or chlorine, for instance as flame retardants \cite{ZHANG20031517}, and polyvinyl chloride consists of more than 50\% chlorine by weight.
Plastics degrade under exposure to ultraviolet (UV) light and can release halogen-containing gases \cite{Khaled2018}.
Their large surface area compared to their volume, the horizontal orientation of their maximum projection area, and long residence times of possibly years, make a complete disintegration of halogen-containing fibers exposed to the extreme UV levels in the stratosphere seem plausible.
The released bromine and chlorine compounds could participate in the catalytic destruction of ozone, similar to those released from the chlorofluorocarbons and halons regulated by the Montreal Protocol.

We have shown that the shape of microplastics is an important factor for their global presence in the environment: the more non-spherical their shape, the larger their horizontal and vertical transport range.
Our findings demonstrate that microplastics can be transported in the atmosphere to almost any point of the globe and are present throughout the troposphere and possibly the stratosphere.
Novel laboratory experiments and model analyses carried out here diminish uncertainties regarding the settling behavior of fibers in the atmosphere, leading to more accurate model simulations of atmospheric concentration and deposition patterns.
The slow settling velocities obtained for fibers make it also seem plausible that microplastic fibers can reach the stratosphere, where they might endanger the ozone layer.
Moreover, particles of other shapes than straight or bent fibers, such as films \cite{GONZALEZPLEITER2020,ALLEN2019} or particles with non-smooth surface textures, may have an even larger atmospheric transport potential.
However, currently both the shape and size distribution of nano- and microplastic particles are largely unknown and more research is needed for their characterization.

\subsection{Acknowledgements}

We thank Constanza Bonadonna for establishing contacts between our groups.
We acknowledge support from the University of Vienna's research platform PLENTY (Plastics in the Environment and Society).
TB was funded by the German Research Foundation (DFG) Walter Benjamin Position (project no.~463393443).
JG was supported by funding from the European Union Horizon 2020 Research and Innovation Programme under the Marie Sklodowska-Curie Actions, Grant Agreement No.675675.
LB was funded by the Austrian Science Fund in the framework of the project P 34170-N, "A demonstration of a Lagrangian re-analysis (LARA)". We further thank Katharina Baier, Marina D\"utsch, Blaz Gasparini, Lukas Kugler, Andreas Plach, Johannes Seezing, Andrey Skorokhod, and Martin Vojta for their support.

\bibliography{doc/latex/achemso/achemso-demo}

\providecommand{\latin}[1]{#1}
\makeatletter
\providecommand{\doi}
  {\begingroup\let\do\@makeother\dospecials
  \catcode`\{=1 \catcode`\}=2 \doi@aux}
\providecommand{\doi@aux}[1]{\endgroup\texttt{#1}}
\makeatother
\providecommand*\mcitethebibliography{\thebibliography}
\csname @ifundefined\endcsname{endmcitethebibliography}
  {\let\endmcitethebibliography\endthebibliography}{}
\begin{mcitethebibliography}{43}
\providecommand*\natexlab[1]{#1}
\providecommand*\mciteSetBstSublistMode[1]{}
\providecommand*\mciteSetBstMaxWidthForm[2]{}
\providecommand*\mciteBstWouldAddEndPuncttrue
  {\def\EndOfBibitem{\unskip.}}
\providecommand*\mciteBstWouldAddEndPunctfalse
  {\let\EndOfBibitem\relax}
\providecommand*\mciteSetBstMidEndSepPunct[3]{}
\providecommand*\mciteSetBstSublistLabelBeginEnd[3]{}
\providecommand*\EndOfBibitem{}
\mciteSetBstSublistMode{f}
\mciteSetBstMaxWidthForm{subitem}{(\alph{mcitesubitemcount})}
\mciteSetBstSublistLabelBeginEnd
  {\mcitemaxwidthsubitemform\space}
  {\relax}
  {\relax}

\bibitem[Seinfeld and Pandis(2016)Seinfeld, and Pandis]{SeinfeldPandis2016}
Seinfeld,~J.~H.; Pandis,~S.~N. \emph{Atmospheric chemistry and physics : from
  air pollution to climate change}, third edition ed.; Wiley: Hoboken, New
  Jersey, 2016\relax
\mciteBstWouldAddEndPuncttrue
\mciteSetBstMidEndSepPunct{\mcitedefaultmidpunct}
{\mcitedefaultendpunct}{\mcitedefaultseppunct}\relax
\EndOfBibitem
\bibitem[Matsui \latin{et~al.}(2011)Matsui, Kondo, Moteki, Takegawa, Sahu,
  Koike, Zhao, Fuelberg, Sessions, Diskin, Anderson, Blake, Wisthaler, Cubison,
  and Jimenez]{Matsui2011}
Matsui,~H.; Kondo,~Y.; Moteki,~N.; Takegawa,~N.; Sahu,~L.~K.; Koike,~M.;
  Zhao,~Y.; Fuelberg,~H.~E.; Sessions,~W.~R.; Diskin,~G.; Anderson,~B.~E.;
  Blake,~D.~R.; Wisthaler,~A.; Cubison,~M.~J.; Jimenez,~J.~L. Accumulation-mode
  aerosol number concentrations in the Arctic during the ARCTAS aircraft
  campaign: Long-range transport of polluted and clean air from the Asian
  continent. \emph{Journal of Geophysical Research: Atmospheres} \textbf{2011},
  \emph{116}\relax
\mciteBstWouldAddEndPuncttrue
\mciteSetBstMidEndSepPunct{\mcitedefaultmidpunct}
{\mcitedefaultendpunct}{\mcitedefaultseppunct}\relax
\EndOfBibitem
\bibitem[Damoah \latin{et~al.}(2004)Damoah, Spichtinger, Forster, James,
  Mattis, Wandinger, Beirle, Wagner, and Stohl]{Damoah2004}
Damoah,~R.; Spichtinger,~N.; Forster,~C.; James,~P.; Mattis,~I.; Wandinger,~U.;
  Beirle,~S.; Wagner,~T.; Stohl,~A. Around the world in 17 days -
  hemispheric-scale transport of forest fire smoke from Russia in May 2003.
  \emph{Atmospheric Chemistry and Physics} \textbf{2004}, \emph{4},
  1311--1321\relax
\mciteBstWouldAddEndPuncttrue
\mciteSetBstMidEndSepPunct{\mcitedefaultmidpunct}
{\mcitedefaultendpunct}{\mcitedefaultseppunct}\relax
\EndOfBibitem
\bibitem[Varga \latin{et~al.}(2021)Varga, Dagsson-Waldhauserov{\'a}, Gresina,
  and Helgadottir]{Varga2021}
Varga,~G.; Dagsson-Waldhauserov{\'a},~P.; Gresina,~F.; Helgadottir,~A. Saharan
  dust and giant quartz particle transport towards Iceland. \emph{Scientific
  Reports} \textbf{2021}, \emph{11}, 11891\relax
\mciteBstWouldAddEndPuncttrue
\mciteSetBstMidEndSepPunct{\mcitedefaultmidpunct}
{\mcitedefaultendpunct}{\mcitedefaultseppunct}\relax
\EndOfBibitem
\bibitem[van~der Does \latin{et~al.}(2018)van~der Does, Knippertz,
  Zschenderlein, Harrison, and Stuut]{Does2018}
van~der Does,~M.; Knippertz,~P.; Zschenderlein,~P.; Harrison,~R.~G.;
  Stuut,~J.-B.~W. The mysterious long-range transport of giant mineral dust
  particles. \emph{Science Advances} \textbf{2018}, \emph{4}, eaau2768\relax
\mciteBstWouldAddEndPuncttrue
\mciteSetBstMidEndSepPunct{\mcitedefaultmidpunct}
{\mcitedefaultendpunct}{\mcitedefaultseppunct}\relax
\EndOfBibitem
\bibitem[Jeong \latin{et~al.}(2014)Jeong, Kim, Seo, Kim, Jin, and
  Chun]{Jeong2013}
Jeong,~G.~Y.; Kim,~J.~Y.; Seo,~J.; Kim,~G.~M.; Jin,~H.~C.; Chun,~Y. Long-range
  transport of giant particles in Asian dust identified by physical,
  mineralogical, and meteorological analysis. \emph{Atmospheric Chemistry and
  Physics} \textbf{2014}, \emph{14}, 505--521\relax
\mciteBstWouldAddEndPuncttrue
\mciteSetBstMidEndSepPunct{\mcitedefaultmidpunct}
{\mcitedefaultendpunct}{\mcitedefaultseppunct}\relax
\EndOfBibitem
\bibitem[Madonna \latin{et~al.}(2010)Madonna, Amodeo, D'Amico, Mona, and
  Pappalardo]{Madonna2010}
Madonna,~F.; Amodeo,~A.; D'Amico,~G.; Mona,~L.; Pappalardo,~G. Observation of
  non-spherical ultragiant aerosol using a microwave radar. \emph{Geophysical
  Research Letters} \textbf{2010}, \emph{37}\relax
\mciteBstWouldAddEndPuncttrue
\mciteSetBstMidEndSepPunct{\mcitedefaultmidpunct}
{\mcitedefaultendpunct}{\mcitedefaultseppunct}\relax
\EndOfBibitem
\bibitem[Middleton \latin{et~al.}(2001)Middleton, Betzer, and
  Bull]{MIDDLETON2001411}
Middleton,~N.; Betzer,~P.; Bull,~P. Long-range transport of ‘giant’ aeolian
  quartz grains: linkage with discrete sedimentary sources and implications for
  protective particle transfer. \emph{Marine Geology} \textbf{2001},
  \emph{177}, 411--417\relax
\mciteBstWouldAddEndPuncttrue
\mciteSetBstMidEndSepPunct{\mcitedefaultmidpunct}
{\mcitedefaultendpunct}{\mcitedefaultseppunct}\relax
\EndOfBibitem
\bibitem[Betzer \latin{et~al.}(1988)Betzer, Carder, Duce, Merrill, Tindale,
  Uematsu, Costello, Young, Feely, Breland, Bernstein, and Greco]{Betzer1988}
Betzer,~P.~R.; Carder,~K.~L.; Duce,~R.~A.; Merrill,~J.~T.; Tindale,~N.~W.;
  Uematsu,~M.; Costello,~D.~K.; Young,~R.~W.; Feely,~R.~A.; Breland,~J.~A.;
  Bernstein,~R.~E.; Greco,~A.~M. Long--range transport of giant mineral aerosol
  particles. \emph{Nature} \textbf{1988}, \emph{336}, 568--571\relax
\mciteBstWouldAddEndPuncttrue
\mciteSetBstMidEndSepPunct{\mcitedefaultmidpunct}
{\mcitedefaultendpunct}{\mcitedefaultseppunct}\relax
\EndOfBibitem
\bibitem[Zhang \latin{et~al.}(2019)Zhang, Sharratt, Lei, Wu, Zhang, Zhao, Wang,
  Wu, Li, Liu, Huang, Guo, Mao, Li, Tang, and Hao]{ZHANG20191}
Zhang,~X.-X. \latin{et~al.}  Parameterization schemes on dust deposition in
  northwest China: Model validation and implications for the global dust cycle.
  \emph{Atmospheric Environment} \textbf{2019}, \emph{209}, 1--13\relax
\mciteBstWouldAddEndPuncttrue
\mciteSetBstMidEndSepPunct{\mcitedefaultmidpunct}
{\mcitedefaultendpunct}{\mcitedefaultseppunct}\relax
\EndOfBibitem
\bibitem[Liu \latin{et~al.}(2014)Liu, Cashman, Beckett, Witham, Leadbetter,
  Hort, and Guomundsson]{Liu2014}
Liu,~E.~J.; Cashman,~K.~V.; Beckett,~F.~M.; Witham,~C.~S.; Leadbetter,~S.~J.;
  Hort,~M.~C.; Guomundsson,~S. Ash mists and brown snow: Remobilization of
  volcanic ash from recent Icelandic eruptions. \emph{Journal of Geophysical
  Research: Atmospheres} \textbf{2014}, \emph{119}, 9463--9480\relax
\mciteBstWouldAddEndPuncttrue
\mciteSetBstMidEndSepPunct{\mcitedefaultmidpunct}
{\mcitedefaultendpunct}{\mcitedefaultseppunct}\relax
\EndOfBibitem
\bibitem[Knippertz \latin{et~al.}(2011)Knippertz, Tesche, Heinold, Kandler,
  Toledano, and Esselborn]{Knippertz2011}
Knippertz,~P.; Tesche,~M.; Heinold,~B.; Kandler,~K.; Toledano,~C.;
  Esselborn,~M. Dust mobilization and aerosol transport from West Africa to
  Cape Verde—a meteorological overview of SAMUM-2. \emph{Chemical and
  Physical Meteorology} \textbf{2011}, \emph{63}, 430--447\relax
\mciteBstWouldAddEndPuncttrue
\mciteSetBstMidEndSepPunct{\mcitedefaultmidpunct}
{\mcitedefaultendpunct}{\mcitedefaultseppunct}\relax
\EndOfBibitem
\bibitem[Garcia-Carreras \latin{et~al.}(2015)Garcia-Carreras, Parker, Marsham,
  Rosenberg, Brooks, Lock, Marenco, McQuaid, and Hobby]{Garcia-Carreras2015}
Garcia-Carreras,~L.; Parker,~D.~J.; Marsham,~J.~H.; Rosenberg,~P.~D.;
  Brooks,~I.~M.; Lock,~A.~P.; Marenco,~F.; McQuaid,~J.~B.; Hobby,~M. The
  Turbulent Structure and Diurnal Growth of the Saharan Atmospheric Boundary
  Layer. \emph{Journal of the Atmospheric Sciences} \textbf{2015}, \emph{72},
  693 -- 713\relax
\mciteBstWouldAddEndPuncttrue
\mciteSetBstMidEndSepPunct{\mcitedefaultmidpunct}
{\mcitedefaultendpunct}{\mcitedefaultseppunct}\relax
\EndOfBibitem
\bibitem[Nicoll(2012)]{Nicoll2012}
Nicoll,~K.~A. Measurements of Atmospheric Electricity Aloft. \emph{Surveys in
  Geophysics} \textbf{2012}, \emph{33}, 991--1057\relax
\mciteBstWouldAddEndPuncttrue
\mciteSetBstMidEndSepPunct{\mcitedefaultmidpunct}
{\mcitedefaultendpunct}{\mcitedefaultseppunct}\relax
\EndOfBibitem
\bibitem[Renard \latin{et~al.}(2018)Renard, Dulac, Durand, Bourgeois, Denjean,
  Vignelles, Cout\'e, Jeannot, Verdier, and Mallet]{Renard2018}
Renard,~J.-B.; Dulac,~F.; Durand,~P.; Bourgeois,~Q.; Denjean,~C.;
  Vignelles,~D.; Cout\'e,~B.; Jeannot,~M.; Verdier,~N.; Mallet,~M. In situ
  measurements of desert dust particles above the western Mediterranean Sea
  with the balloon-borne Light Optical Aerosol Counter/sizer (LOAC) during the
  ChArMEx campaign of summer 2013. \emph{Atmospheric Chemistry and Physics}
  \textbf{2018}, \emph{18}, 3677--3699\relax
\mciteBstWouldAddEndPuncttrue
\mciteSetBstMidEndSepPunct{\mcitedefaultmidpunct}
{\mcitedefaultendpunct}{\mcitedefaultseppunct}\relax
\EndOfBibitem
\bibitem[Saxby \latin{et~al.}(2018)Saxby, Beckett, Cashman, Rust, and
  Tennant]{SAXBY2018}
Saxby,~J.; Beckett,~F.; Cashman,~K.; Rust,~A.; Tennant,~E. The impact of
  particle shape on fall velocity: Implications for volcanic ash dispersion
  modelling. \emph{Journal of Volcanology and Geothermal Research}
  \textbf{2018}, \emph{362}, 32--48\relax
\mciteBstWouldAddEndPuncttrue
\mciteSetBstMidEndSepPunct{\mcitedefaultmidpunct}
{\mcitedefaultendpunct}{\mcitedefaultseppunct}\relax
\EndOfBibitem
\bibitem[Bergmann \latin{et~al.}(2019)Bergmann, Mützel, Primpke, Tekman,
  Trachsel, and Gerdts]{BERGMANN2019}
Bergmann,~M.; Mützel,~S.; Primpke,~S.; Tekman,~M.~B.; Trachsel,~J.; Gerdts,~G.
  White and wonderful? Microplastics prevail in snow from the Alps to the
  Arctic. \emph{Science Advances} \textbf{2019}, \emph{5}, eaax1157\relax
\mciteBstWouldAddEndPuncttrue
\mciteSetBstMidEndSepPunct{\mcitedefaultmidpunct}
{\mcitedefaultendpunct}{\mcitedefaultseppunct}\relax
\EndOfBibitem
\bibitem[Allen \latin{et~al.}(2019)Allen, Allen, Phoenix, Roux, Jiménez,
  Simonneau, Binet, and Galop]{ALLEN2019}
Allen,~S.; Allen,~D.; Phoenix,~V.~R.; Roux,~G.~L.; Jiménez,~P.~D.;
  Simonneau,~A.; Binet,~S.; Galop,~D. Atmospheric transport and deposition of
  microplastics in a remote mountain catchment. \emph{Nature Geoscience}
  \textbf{2019}, \emph{5}\relax
\mciteBstWouldAddEndPuncttrue
\mciteSetBstMidEndSepPunct{\mcitedefaultmidpunct}
{\mcitedefaultendpunct}{\mcitedefaultseppunct}\relax
\EndOfBibitem
\bibitem[Ambrosini \latin{et~al.}(2019)Ambrosini, Azzoni, Pittino, Diolaiuti,
  Franzetti, and Parolini]{AMBROSINI2019}
Ambrosini,~R.; Azzoni,~R.~S.; Pittino,~F.; Diolaiuti,~G.; Franzetti,~A.;
  Parolini,~M. First evidence of microplastic contamination in the supraglacial
  debris of an alpine glacier. \emph{Environmental Pollution} \textbf{2019},
  \emph{253}, 297--301\relax
\mciteBstWouldAddEndPuncttrue
\mciteSetBstMidEndSepPunct{\mcitedefaultmidpunct}
{\mcitedefaultendpunct}{\mcitedefaultseppunct}\relax
\EndOfBibitem
\bibitem[Brahney \latin{et~al.}(2021)Brahney, Mahowald, Prank, Cornwell,
  Klimont, Matsui, and Prather]{Brahney2021}
Brahney,~J.; Mahowald,~N.; Prank,~M.; Cornwell,~G.; Klimont,~Z.; Matsui,~H.;
  Prather,~K.~A. Constraining the atmospheric limb of the plastic cycle.
  \emph{Proceedings of the National Academy of Sciences} \textbf{2021},
  \emph{118}, e2020719118\relax
\mciteBstWouldAddEndPuncttrue
\mciteSetBstMidEndSepPunct{\mcitedefaultmidpunct}
{\mcitedefaultendpunct}{\mcitedefaultseppunct}\relax
\EndOfBibitem
\bibitem[Stefánsson \latin{et~al.}(2021)Stefánsson, Peternell,
  Konrad-Schmolke, Hannesdóttir, Ásbjörnsson, and Sturkell]{Stefansson2021}
Stefánsson,~H.; Peternell,~M.; Konrad-Schmolke,~M.; Hannesdóttir,~H.;
  Ásbjörnsson,~E.~J.; Sturkell,~E. Microplastics in Glaciers: First Results
  from the Vatnajökull Ice Cap. \emph{Sustainability} \textbf{2021},
  \emph{13}\relax
\mciteBstWouldAddEndPuncttrue
\mciteSetBstMidEndSepPunct{\mcitedefaultmidpunct}
{\mcitedefaultendpunct}{\mcitedefaultseppunct}\relax
\EndOfBibitem
\bibitem[Cabrera \latin{et~al.}(2020)Cabrera, Valencia, Lucas-Solis, Calero,
  Maisincho, Conicelli, {Massaine Moulatlet}, and Capparelli]{CABRERA2020}
Cabrera,~M.; Valencia,~B.~G.; Lucas-Solis,~O.; Calero,~J.~L.; Maisincho,~L.;
  Conicelli,~B.; {Massaine Moulatlet},~G.; Capparelli,~M.~V. A new method for
  microplastic sampling and isolation in mountain glaciers: A case study of one
  antisana glacier, Ecuadorian Andes. \emph{Case Studies in Chemical and
  Environmental Engineering} \textbf{2020}, \emph{2}, 100051\relax
\mciteBstWouldAddEndPuncttrue
\mciteSetBstMidEndSepPunct{\mcitedefaultmidpunct}
{\mcitedefaultendpunct}{\mcitedefaultseppunct}\relax
\EndOfBibitem
\bibitem[González-Pleiter \latin{et~al.}(2020)González-Pleiter, Edo,
  Velázquez, Casero-Chamorro, Leganés, Quesada, Fernández-Piñas, and
  Rosal]{GONZALEZPLEITER2020}
González-Pleiter,~M.; Edo,~C.; Velázquez,~D.; Casero-Chamorro,~M.~C.;
  Leganés,~F.; Quesada,~A.; Fernández-Piñas,~F.; Rosal,~R. First detection
  of microplastics in the freshwater of an Antarctic Specially Protected Area.
  \emph{Marine Pollution Bulletin} \textbf{2020}, \emph{161}, 111811\relax
\mciteBstWouldAddEndPuncttrue
\mciteSetBstMidEndSepPunct{\mcitedefaultmidpunct}
{\mcitedefaultendpunct}{\mcitedefaultseppunct}\relax
\EndOfBibitem
\bibitem[Bagheri and Bonadonna(2016)Bagheri, and Bonadonna]{BB2016}
Bagheri,~G.; Bonadonna,~C. On the drag of freely falling non-spherical
  particles. \emph{Powder Technology} \textbf{2016}, 526--544\relax
\mciteBstWouldAddEndPuncttrue
\mciteSetBstMidEndSepPunct{\mcitedefaultmidpunct}
{\mcitedefaultendpunct}{\mcitedefaultseppunct}\relax
\EndOfBibitem
\bibitem[Mallios \latin{et~al.}(2020)Mallios, Drakaki, and
  Amiridis]{Mallios2020}
Mallios,~S.~A.; Drakaki,~E.; Amiridis,~V. Effects of dust particle sphericity
  and orientation on their gravitational settling in the earth’s atmosphere.
  \emph{Journal of Aerosol Science} \textbf{2020}, \emph{150}, 105634\relax
\mciteBstWouldAddEndPuncttrue
\mciteSetBstMidEndSepPunct{\mcitedefaultmidpunct}
{\mcitedefaultendpunct}{\mcitedefaultseppunct}\relax
\EndOfBibitem
\bibitem[Bhowmick \latin{et~al.}(2023)Bhowmick, Seesing, Gustavsson, Guettler,
  Pumir, Mehlig, Wang, and Bagheri]{Bhowmick2023}
Bhowmick,~T.; Seesing,~J.; Gustavsson,~K.; Guettler,~J.; Pumir,~A.; Mehlig,~B.;
  Wang,~Y.; Bagheri,~G. Inertial angular dynamics of non-spherical atmospheric
  particles. \emph{arXiv preprint arXiv:2303.04299} \textbf{2023}, \relax
\mciteBstWouldAddEndPunctfalse
\mciteSetBstMidEndSepPunct{\mcitedefaultmidpunct}
{}{\mcitedefaultseppunct}\relax
\EndOfBibitem
\bibitem[Newsom and Bruce(1994)Newsom, and Bruce]{Newsom1994}
Newsom,~R.~K.; Bruce,~C.~W. The dynamics of fibrous aerosols in a quiescent
  atmosphere. \emph{Physics of Fluids} \textbf{1994}, \emph{6}, 521--530\relax
\mciteBstWouldAddEndPuncttrue
\mciteSetBstMidEndSepPunct{\mcitedefaultmidpunct}
{\mcitedefaultendpunct}{\mcitedefaultseppunct}\relax
\EndOfBibitem
\bibitem[Jayaweera and Cottis(1969)Jayaweera, and Cottis]{JayaweeraCottis1969}
Jayaweera,~K. O. L.~F.; Cottis,~R.~E. Fall velocities of plate-like and
  columnar ice crystals. \emph{Quarterly Journal of the Royal Meteorological
  Society} \textbf{1969}, \emph{95}, 703--709\relax
\mciteBstWouldAddEndPuncttrue
\mciteSetBstMidEndSepPunct{\mcitedefaultmidpunct}
{\mcitedefaultendpunct}{\mcitedefaultseppunct}\relax
\EndOfBibitem
\bibitem[scr()]{NanoScribe_2017}
Photonic Professional (GT) User Manual. Nanoscribe GmbH\relax
\mciteBstWouldAddEndPuncttrue
\mciteSetBstMidEndSepPunct{\mcitedefaultmidpunct}
{\mcitedefaultendpunct}{\mcitedefaultseppunct}\relax
\EndOfBibitem
\bibitem[Qi \latin{et~al.}(2012)Qi, Nathan, and Kelso]{QI020121}
Qi,~G.~Q.; Nathan,~G.~J.; Kelso,~R.~M. PTV measurement of drag coefficient of
  fibrous particles with large aspect ratio. \emph{Powder Technology}
  \textbf{2012}, \emph{229}, 261--269\relax
\mciteBstWouldAddEndPuncttrue
\mciteSetBstMidEndSepPunct{\mcitedefaultmidpunct}
{\mcitedefaultendpunct}{\mcitedefaultseppunct}\relax
\EndOfBibitem
\bibitem[Coyle \latin{et~al.}(2023)Coyle, Service, Witte, Hardiman, and
  McKinley]{Coyle2023}
Coyle,~R.; Service,~M.; Witte,~U.; Hardiman,~G.; McKinley,~J. Modeling
  Microplastic Transport in the Marine Environment: Testing Empirical Models of
  Particle Terminal Sinking Velocity for Irregularly Shaped Particles.
  \emph{ACS ES\&T Water} \textbf{2023}, \relax
\mciteBstWouldAddEndPunctfalse
\mciteSetBstMidEndSepPunct{\mcitedefaultmidpunct}
{}{\mcitedefaultseppunct}\relax
\EndOfBibitem
\bibitem[Stohl \latin{et~al.}(2005)Stohl, Forster, Frank, Seibert, and
  Wotawa]{Stohl2005}
Stohl,~A.; Forster,~C.; Frank,~A.; Seibert,~P.; Wotawa,~G. Technical note: The
  Lagrangian particle dispersion model FLEXPART version 6.2. \emph{Atmospheric
  Chemistry and Physics} \textbf{2005}, \emph{5}, 2461--2474\relax
\mciteBstWouldAddEndPuncttrue
\mciteSetBstMidEndSepPunct{\mcitedefaultmidpunct}
{\mcitedefaultendpunct}{\mcitedefaultseppunct}\relax
\EndOfBibitem
\bibitem[Pisso \latin{et~al.}(2019)Pisso, Sollum, Grythe, Kristiansen,
  Cassiani, Eckhardt, Arnold, Morton, Thompson, Groot~Zwaaftink, Evangeliou,
  Sodemann, Haimberger, Henne, Brunner, Burkhart, Fouilloux, Brioude, Philipp,
  Seibert, and Stohl]{Pisso2019}
Pisso,~I. \latin{et~al.}  The Lagrangian particle dispersion model FLEXPART
  version 10.4. \emph{Geoscientific Model Development} \textbf{2019},
  \emph{12}, 4955--4997\relax
\mciteBstWouldAddEndPuncttrue
\mciteSetBstMidEndSepPunct{\mcitedefaultmidpunct}
{\mcitedefaultendpunct}{\mcitedefaultseppunct}\relax
\EndOfBibitem
\bibitem[Hamilton \latin{et~al.}(2021)Hamilton, Bourdages, Geoffroy, Vermaire,
  Mallory, Rochman, and Provencher]{HAMILTON2021}
Hamilton,~B.~M.; Bourdages,~M.~P.; Geoffroy,~C.; Vermaire,~J.~C.;
  Mallory,~M.~L.; Rochman,~C.~M.; Provencher,~J.~F. Microplastics around an
  Arctic seabird colony: Particle community composition varies across
  environmental matrices. \emph{Science of The Total Environment}
  \textbf{2021}, \emph{773}, 145536\relax
\mciteBstWouldAddEndPuncttrue
\mciteSetBstMidEndSepPunct{\mcitedefaultmidpunct}
{\mcitedefaultendpunct}{\mcitedefaultseppunct}\relax
\EndOfBibitem
\bibitem[Lebreton \latin{et~al.}(2017)Lebreton, van~der Zwet~J, Damsteeg, Slat,
  Andrady, and Reisser]{Lebreton2017}
Lebreton,~L.~C.; van~der Zwet~J,~J.; Damsteeg,~J.-W.; Slat,~B.; Andrady,~A.;
  Reisser,~J. River plastic emissions to the world’s oceans. \emph{Nature
  Communications} \textbf{2017}, \emph{8}, 15611\relax
\mciteBstWouldAddEndPuncttrue
\mciteSetBstMidEndSepPunct{\mcitedefaultmidpunct}
{\mcitedefaultendpunct}{\mcitedefaultseppunct}\relax
\EndOfBibitem
\bibitem[Jambeck \latin{et~al.}(2015)Jambeck, Geyer, Wilcox, Siegler, Perryman,
  Andrady, Narayan, and Law]{Jambeck2015}
Jambeck,~J.~R.; Geyer,~R.; Wilcox,~C.; Siegler,~T.~R.; Perryman,~M.;
  Andrady,~A.; Narayan,~R.; Law,~K.~L. Plastic waste inputs from land into the
  ocean. \emph{Science} \textbf{2015}, \emph{347}, 768--771\relax
\mciteBstWouldAddEndPuncttrue
\mciteSetBstMidEndSepPunct{\mcitedefaultmidpunct}
{\mcitedefaultendpunct}{\mcitedefaultseppunct}\relax
\EndOfBibitem
\bibitem[Andrady(2011)]{ANDRADY2011}
Andrady,~A.~L. Microplastics in the marine environment. \emph{Marine Pollution
  Bulletin} \textbf{2011}, \emph{62}, 1596--1605\relax
\mciteBstWouldAddEndPuncttrue
\mciteSetBstMidEndSepPunct{\mcitedefaultmidpunct}
{\mcitedefaultendpunct}{\mcitedefaultseppunct}\relax
\EndOfBibitem
\bibitem[Ganguly and Ariya(2019)Ganguly, and Ariya]{Ganguly2019}
Ganguly,~M.; Ariya,~P.~A. Ice Nucleation of Model Nanoplastics and
  Microplastics: A Novel Synthetic Protocol and the Influence of Particle
  Capping at Diverse Atmospheric Environments. \emph{ACS Earth and Space
  Chemistry} \textbf{2019}, \emph{3}, 1729--1739\relax
\mciteBstWouldAddEndPuncttrue
\mciteSetBstMidEndSepPunct{\mcitedefaultmidpunct}
{\mcitedefaultendpunct}{\mcitedefaultseppunct}\relax
\EndOfBibitem
\bibitem[Diallo \latin{et~al.}(2021)Diallo, Ern, and Ploeger]{Diallo2021}
Diallo,~M.; Ern,~M.; Ploeger,~F. The advective Brewer-Dobson circulation in the
  ERA5 reanalysis: climatology, variability, and trends. \emph{Atmospheric
  Chemistry and Physics} \textbf{2021}, \emph{21}, 7515--7544\relax
\mciteBstWouldAddEndPuncttrue
\mciteSetBstMidEndSepPunct{\mcitedefaultmidpunct}
{\mcitedefaultendpunct}{\mcitedefaultseppunct}\relax
\EndOfBibitem
\bibitem[Flury \latin{et~al.}(2013)Flury, Wu, and Read]{Flury2013}
Flury,~T.; Wu,~D.~L.; Read,~W.~G. Variability in the speed of the
  Brewer–Dobson circulation as observed by Aura/MLS. \emph{Atmospheric
  Chemistry and Physics} \textbf{2013}, \emph{13}, 4563--4575\relax
\mciteBstWouldAddEndPuncttrue
\mciteSetBstMidEndSepPunct{\mcitedefaultmidpunct}
{\mcitedefaultendpunct}{\mcitedefaultseppunct}\relax
\EndOfBibitem
\bibitem[Zhang and Horrocks(2003)Zhang, and Horrocks]{ZHANG20031517}
Zhang,~S.; Horrocks,~A. A review of flame retardant polypropylene fibres.
  \emph{Progress in Polymer Science} \textbf{2003}, \emph{28}, 1517--1538\relax
\mciteBstWouldAddEndPuncttrue
\mciteSetBstMidEndSepPunct{\mcitedefaultmidpunct}
{\mcitedefaultendpunct}{\mcitedefaultseppunct}\relax
\EndOfBibitem
\bibitem[Khaled \latin{et~al.}(2018)Khaled, Rivaton, Richard, Jaber, and
  Sleiman]{Khaled2018}
Khaled,~A.; Rivaton,~A.; Richard,~C.; Jaber,~F.; Sleiman,~M.
  Phototransformation of Plastic Containing Brominated Flame Retardants:
  Enhanced Fragmentation and Release of Photoproducts to Water and Air.
  \emph{Environmental Science {\&} Technology} \textbf{2018}, \emph{52},
  11123--11131\relax
\mciteBstWouldAddEndPuncttrue
\mciteSetBstMidEndSepPunct{\mcitedefaultmidpunct}
{\mcitedefaultendpunct}{\mcitedefaultseppunct}\relax
\EndOfBibitem
\end{mcitethebibliography}

\end{document}